\documentclass[twocolumn,showpacs,preprintnumbers,amsmath,amssymb,prb]{revtex4}

\usepackage{graphicx}
\usepackage{dcolumn}
\usepackage{bm}
\usepackage{subfigure}
\usepackage{amsmath}
\usepackage{color}

\begingroup
  \catcode`\_=\active
  \gdef_#1{\ensuremath{\sb{\mathrm{#1}}}}
\endgroup
\mathcode`\_=\string"8000
\catcode`\_=12

\definecolor{green}{rgb}{0,0.5,0}

\begin{document}

\preprint{APS/123-QED}

\title{Measurement and modeling of large area normal-metal/insulator/superconductor refrigerator with improved cooling\footnote{Contribution of a U.S. government agency, not subject to copyright.}}

\author{Galen C. O'Neil}
\email{oneilg@nist.gov}
\author{Peter J. Lowell}
 \altaffiliation[Also at ]{University of Colorado, Boulder}
\author{Jason M. Underwood}
\author{Joel N. Ullom}%
 \email{ullom@nist.gov}
\affiliation{%
National Institute of Standards and Technology, Boulder, CO
}%

\date{\today}

\begin{abstract}
In a normal-metal/insulator/superconductor (NIS) tunnel junction refrigerator, the normal-metal electrons are cooled and the dissipated power heats the superconducting electrode.  This paper presents a review of the mechanisms by which heat leaves the superconductor and introduces overlayer quasiparticle traps for more effective heatsinking.  A comprehensive thermal model is presented that accounts for the described physics, including the behavior of athermal phonons generated by both quasiparticle recombination and trapped quasiparticles.  We compare the model to measurements of a large area ($>400~\mu m^2$) NIS refrigerator with overlayer quasiparticle traps, and demonstrate that the model is in good agreement experiment.  The refrigerator IV curve at a bath temperature of 300~mK is consistent with an electron temperature of 82~mK.  However, evidence from independent thermometer junctions suggests that the refrigerator junction is creating an athermal electron distribution whose total excitation energy corresponds to a higher temperature than is indicated by the refrigerator IV curve.  \end{abstract}

\maketitle

\begin{figure}
\includegraphics[width=0.9\linewidth]{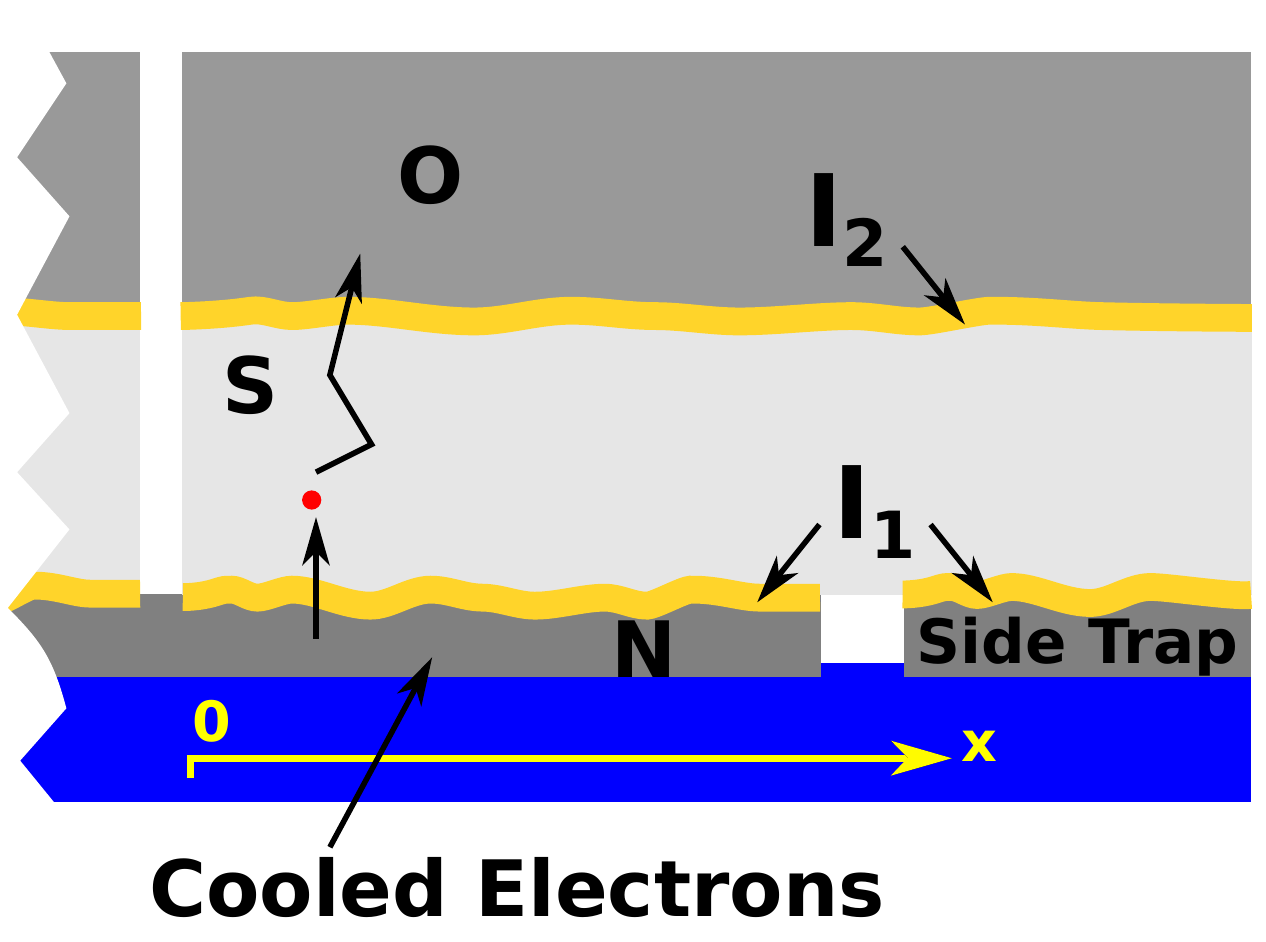}
\caption{\label{fig:junction_sideview} Cross-sectional sketch of an NIS  refrigerator with overlayer trap.  A quasiparticle (red dot) is depicted diffusing through the superconductor to the overlayer trap.  Layers are labeled $O$ for the overlayer trap, $S$ for the superconductor, $N$ for the cooled normal metal layer, I$_1$ and I$_2$ for insulating layers.  Because I$_2$ is fabricated independently of I$_1$, it can have a different transparency. }
\end{figure}

\section{Introduction}

In a normal-metal/insulator/superconductor (NIS) tunnel junction biased near the superconducting gap energy $\Delta$, the single quasiparticle tunneling current transfers heat from normal metal electrons to the superconductor.  This transfer enables refrigerators that can cool electrons from 300~mK to $\sim$100~mK,\cite{Leivo1996, Pekola2004} and can cool arbitrary payloads as well.  For example, NIS refrigerators have been used to cool a macroscopic germanium thermometer\cite{Clark2005} and a superconducting transition edge x-ray detector.\cite{miller2008_apl}  The performance of these refrigerators is limited, in part, by heating of the superconductor due to the dissipated power $IV$, and the heat removed from the normal metal.  The impact of this heating on NIS refrigerator performance is often characterized by the fraction $\beta$ of the power deposited in the superconductor that returns to the normal metal as an excess load.\cite{Ullom2000a}

Previous efforts to model the heating of the superconductor in NIS refrigerators to predict $\beta$ have included quasiparticle diffusion, trapping and recombination.  For example, Ullom and Fisher  numerically solved a differential equation for the excess quasiparticle density vs position in the superconductor.\cite{Ullom2000a}  Rajauria et al\cite{PhysRevB.80.214521} used approximations to solve similar differential equations analytically and introduced a finite quasiparticle trapping rate.  In that work, a parameter equivalent to $\beta$ was calculated, and the agreement with experiment was within a factor of 3 to 10.

In this paper, we expand upon previous work by providing the most comprehensive model of NIS refrigeration to date.  We add a new form of quasiparticle traps, referred to as overlayer traps, to both the model and devices.  We model not only  the superconductor quasiparticle temperature, but also the overlayer trap electron temperature and the temperature of the phonons in the metal layers that make up the NIS refrigerator.   We also account for the athermal behavior of excitations with energy $\gg k_b T$ created by quasiparticle relaxation.  Our implementation of this model easily handles changes in nearly every input parameter, allowing us to examine a large area of parameter space and design the next generation of NIS refrigerators.  We present measurements on a large area NIS refrigerator that agree with the model predictions over a large temperature range. The refrigerator IV curves are consistent with cooling from 300~mK to 82~mK.  As discussed in Sec \ref{sec:athermal} the refrigerator may be creating an athermal electron distribution so the interpretation of the cooling results is not straightforward.

\begin{figure}
\includegraphics[width=0.9\linewidth]{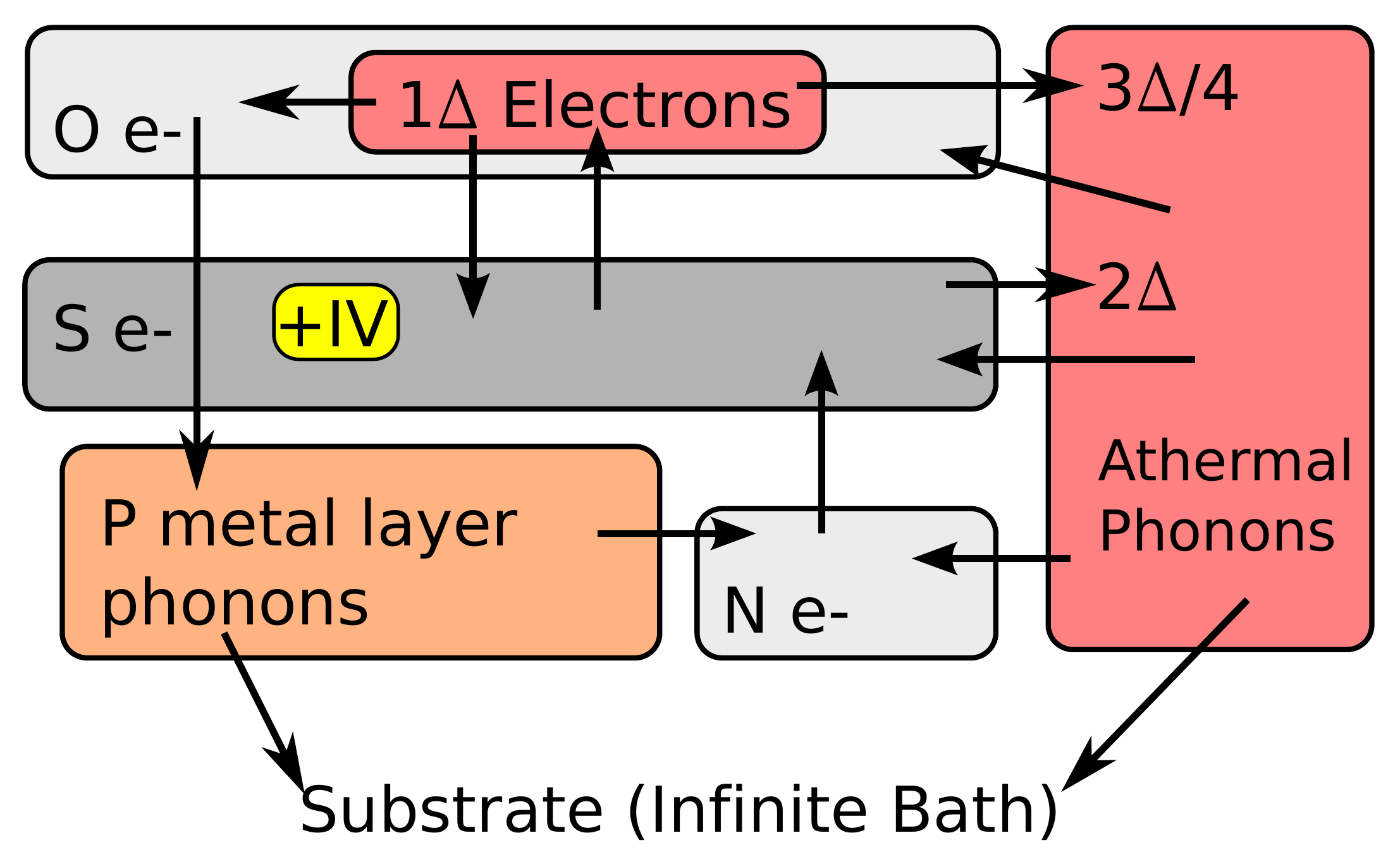}
\caption{\label{fig:thermalmodel} Block diagram representation of the thermal model.  Roughly, power is deposited in the superconductor and leaves by trapping to the overlayer traps.  Power leaves the electrons of the overlayer trap by coupling to phonons in the metal layer, and the phonons couple to the substrate (the bath) via a boundary resistance.  There is additional complexity due to quasiparticle recombination in the superconductor and the behavior of athermal phonons created by recombination and trapped quasiparticles. }
\end{figure}

\section{Model Overview}
\label{sec:model_overview}

We model a single NIS junction, which makes up half of an SINIS refrigerator, as shown in Fig \ref{fig:junction_sideview}.  The important systems in the device are the cooled normal metal electrons $N$, the superconductor quasiparticles $S$, the overlayer electrons $O$, and the combined metal layer phonons $P$.    Roughly speaking, the power flow is as follows: a power $P_S$ is deposited in the $S$ layer by the NIS refrigerator junction, which increases the quasiparticle density locally above the junction.  These quasiparticles may diffuse and recombine, but the majority are trapped into the $O$ layer.  At this point, they become electronic excitations with energy $\Delta \gg k_b T$.  Various processes allow these electrons to relax, and the majority of the energy couples via electron-phonon coupling to the $P$ system.  Because all three metal layers are made of Al, or AlMn, with only thin oxide layers between them, we assume that there is one phonon system shared by all three metal layers.  Finally, the phonons in the metal layers relax by coupling to the substrate phonons.  Systems in the model are connected by both thermal and athermal processes. The model is summarized as a block diagram in Fig \ref{fig:thermalmodel}.

Section \ref{sec:superconductors_and_tunneling} discusses NIS tunneling, Section \ref{sec:qp_trapping} describes quasiparticle trapping, and a detailed discussion of the rest of the physics underlying the model is found in Appendices \ref{appendix1}--\ref{sec:phonon_escape}. Table \ref{table:central_params_300mK} shows all the model parameters and typical values.  Section \ref{sec:experiment} compares the model predictions to measurements on an NIS refrigerator device.


The model consists of four coupled equations, three of which are position dependent. Equation \ref{eq_diff_O} describes the overlayer trap electron temperature $T_O$. The terms on the right hand side from left to right are due to trapping from $S$, coupling to $P$, and recombination phonons.  Equation \ref{eq_diff_S} describes the excess quasiparticle density in the superconductor $n_{ex}$. The terms from left to right are injection by the junction, trapping to $O$, recombination, and trapping to the side-traps.  Equation \ref{eq_diff_P} describes the temperature of the phonons in the combined metal layers $T_P$. The terms from left to right are coupling to $O$, coupling to $N$ layer, and coupling to the substrate. Equation \ref{eq_powerbalance_beta} is a power balance equation for the electron temperature $T_N$ of the $N$ layer. The terms from left to right are due to the tunnel junction ($P_N$ is negative during refrigeration), Joule heating, coupling to $P$, two-particle tunneling (Andreev reflections), recombination phonons, phonons generated by trapped quasiparticles, and either stray power or power from a payload.

\begin{widetext}
\begin{align}
\label{eq_diff_O}
t_O \kappa_O \frac{d^2 T_O}{dx^2} &= -(\Pi_{e}+\frac{\Pi_{p}}{4}+\frac{3 \Pi_{p}}{4}A_{3\Delta/4-O})\mathcal{P}_{trap}+t_O\mathcal{P}_{p-e} -t_S A_{2\Delta-O}\Gamma_{R}(n_{ex}^2+2 n_{ex} n_{th})\Delta\\
\label{eq_diff_S}
t_S D_{S-I}\frac{d^2 n_{ex}}{dx^2} &= -g\mathcal{P}_S \Delta^{-1} + (1-\Pi_{tun})\mathcal{P}_{trap}\Delta^{-1} 
+ (1-A_{2\Delta-S})\Gamma_{R}(n_{ex}^2+2 n_{ex} n_{th}) +g_{side-trap}\mathcal{P}_{side-trap}\Delta^{-1}\\
\label{eq_diff_P}
(t_O+t_s+&t_N)\kappa_p \frac{d^2 T_P}{dx^2} = -t_O \mathcal{P}_{p-e} + g t_N \mathcal{P}_{p-e} +  \mathcal{P}_{amm}\\
0&=P_N+I_{NS}^2 R_{pad}+P_{p-e}+I_2 V_b+ P_{qp-recomb} + P_{trap-phonons} + P_{excess}.
\label{eq_powerbalance_beta}
\end{align}
\end{widetext}
The variables $t_O$, $t_S$, $t_P$, and $t_N$ are the thicknesses of the $O$, $S$, $P$, and $N$ layers, $\kappa_O$ is the electronic thermal conductivity of the $O$ layer (Eq \ref{eq_weidemannfranz}), the $\Pi_x$ variables describe the relaxation branching ratios of trapped quasiparticles (Eq \ref{eq:pi_all}), the $A_{x-y}$ variables describe the probability of absorption of athermal phonons of energy $x$ in layer $y$ (Sec \ref{section_athermalphonons}),  $\Gamma_R(n_{ex}^2+2n_{ex} n_{th})$ is the recombination rate of excess quasiparticles (Eq \ref{eq_excess_qp_recombination}), $n_{th}$ is the thermal quasiparticle density due to the bath temperature, $D_{S-I}$ is the quasiparticle diffusion constant (Eq \ref{eq_qp_diffusion_injection}), $g$ and $g_{side-trap}$ are functions describing the location of the refrigerator junction and side-traps, $\kappa_P$ is the metal layer phonon thermal conductivity (Eq \ref{kappa_p}), $\mathcal{P}_{trap}$ is the quasiparticle trapping rate (Eq \ref{eq_ptrap}), $\mathcal{P}_{p-e}$ and $P_{p-e}$ are electron phonon coupling terms (Eq \ref{ep_epcoupling}), $\mathcal{P}_S$ is the power deposited in the superconductor by the NIS junction (Eq \ref{eq:ps}), and $\mathcal{P}_{amm}$ is power flow across the acoustic mismatch between the Al layers and the substrate (Eq \ref{eq:pamm}). Script $\mathcal{P}$ terms have units power per unit area or volume, where non-script $P$ terms have units of power.  All equations are solved numerically, and Eqs \ref{eq_diff_O}--\ref{eq_diff_P}  have boundary conditions $dT/dx=0$ at both $x=0$ and $x=x_{end}$.\cite{O'Neil2011}

In the power balance equation (Eq \ref{eq_powerbalance_beta}) power $P_N$ is deposited in the $N$ layer by the NIS junction ($P_N$ is negative during refrigeration).  The temperature reduction is limited by an electron-phonon coupling power $P_{p-e}$, Joule heating $I_{NS}^2 R_{pad}$, two-particle tunneling dissipation $I_2 V_b$, incident phonon power from quasiparticle recombination $P_{qp-recomb}=A_{2\Delta-N}t_NA_{NS}\Delta\Gamma_{R}(n_{ex}^2+2 n_{ex} n_{th})$, and incident phonon power from trapped quasiparticles $P_{trap-phonons}=\frac{3\Pi_p}{4}A_{3\Delta/4-N}\mathcal{P}_{trap}A_{NS}$. Here, $R_{pad}$ is the resistance of the current path through the $N$ layer, $I_{NS}$ is the total tunneling current, $V_b$ is the junction bias voltage, and $I_2$ is the two particle  tunneling current  (Eq~\ref{eq:twoparticletunneling}).\cite{PhysRevLett.100.207002}  An additional power $P_{excess}$ may be present, such as stray RF power or power dissipated by a cooled payload.  A term $\beta P_S$ is often included in Eq \ref{eq_powerbalance_beta} and used as an empirical parameter that accounts for heating of the superconductor.  For the purposes of this paper we use $\beta=0$ and account for heating of the superconductor explicitly.  One goal of the comprehensive thermal model is to predict $\beta$ from measurable NIS design and material parameters.  When we compare model and experimental results in Section \ref{sec:experiment}, we use $\beta$ as a comparison metric.  

\begin{figure}
\includegraphics[width=1\linewidth]{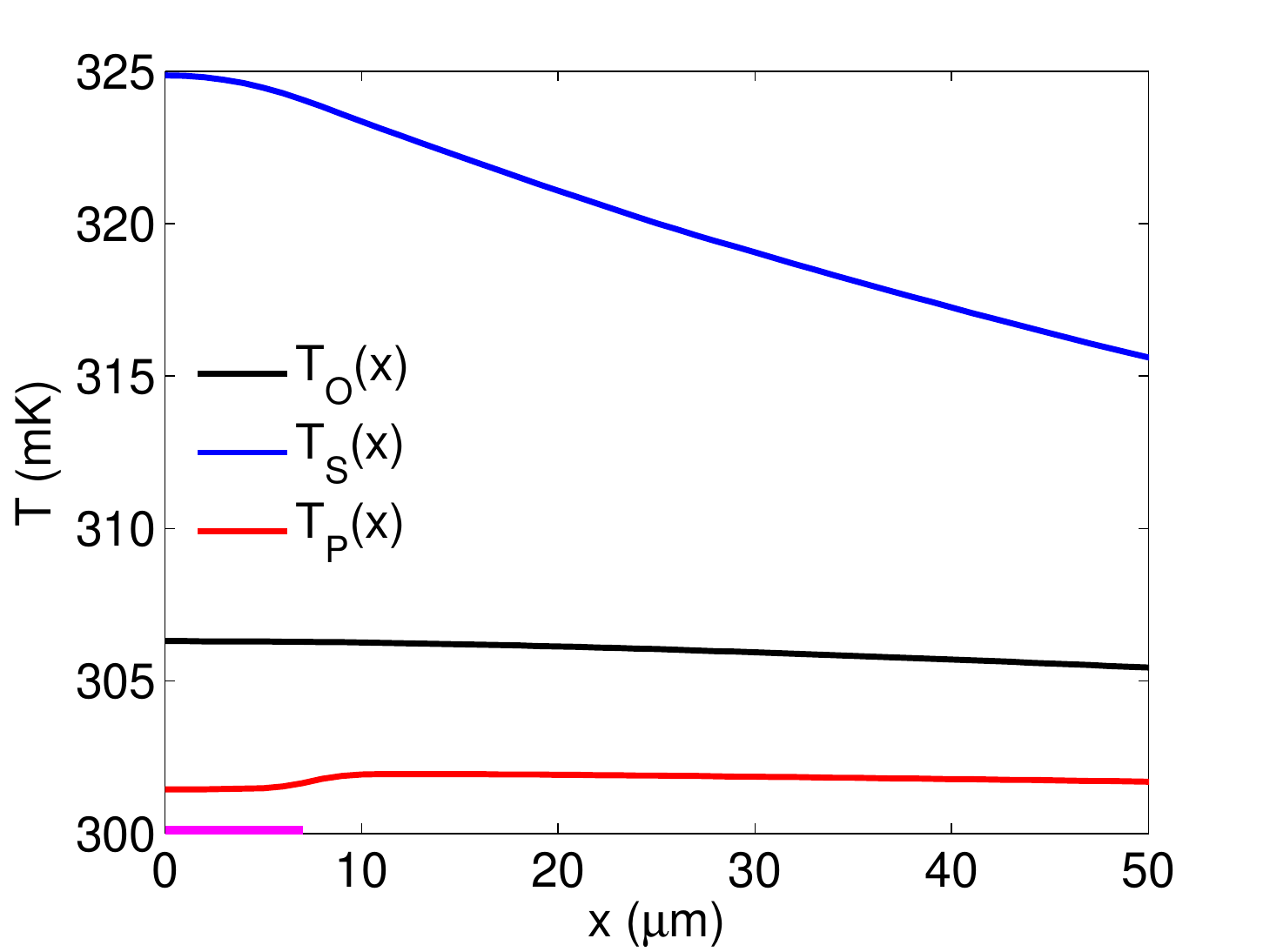}
\caption{\label{fig:modeltemps} Temperatures of the superconductor quasiparticles, overlayer trap electrons and the phonons of the combined metal layers vs position shown from $x=0$ to $x=50~\mu$m calculated with the comprehensive thermal model and parameters from Table \ref{table:central_params_300mK}.  The pink bar on the x-axis shows the location of the refrigerator junction, and therefore, of quasiparticle injection into the superconductor. }
\end{figure}

\begin{figure}
\includegraphics[width=1\linewidth]{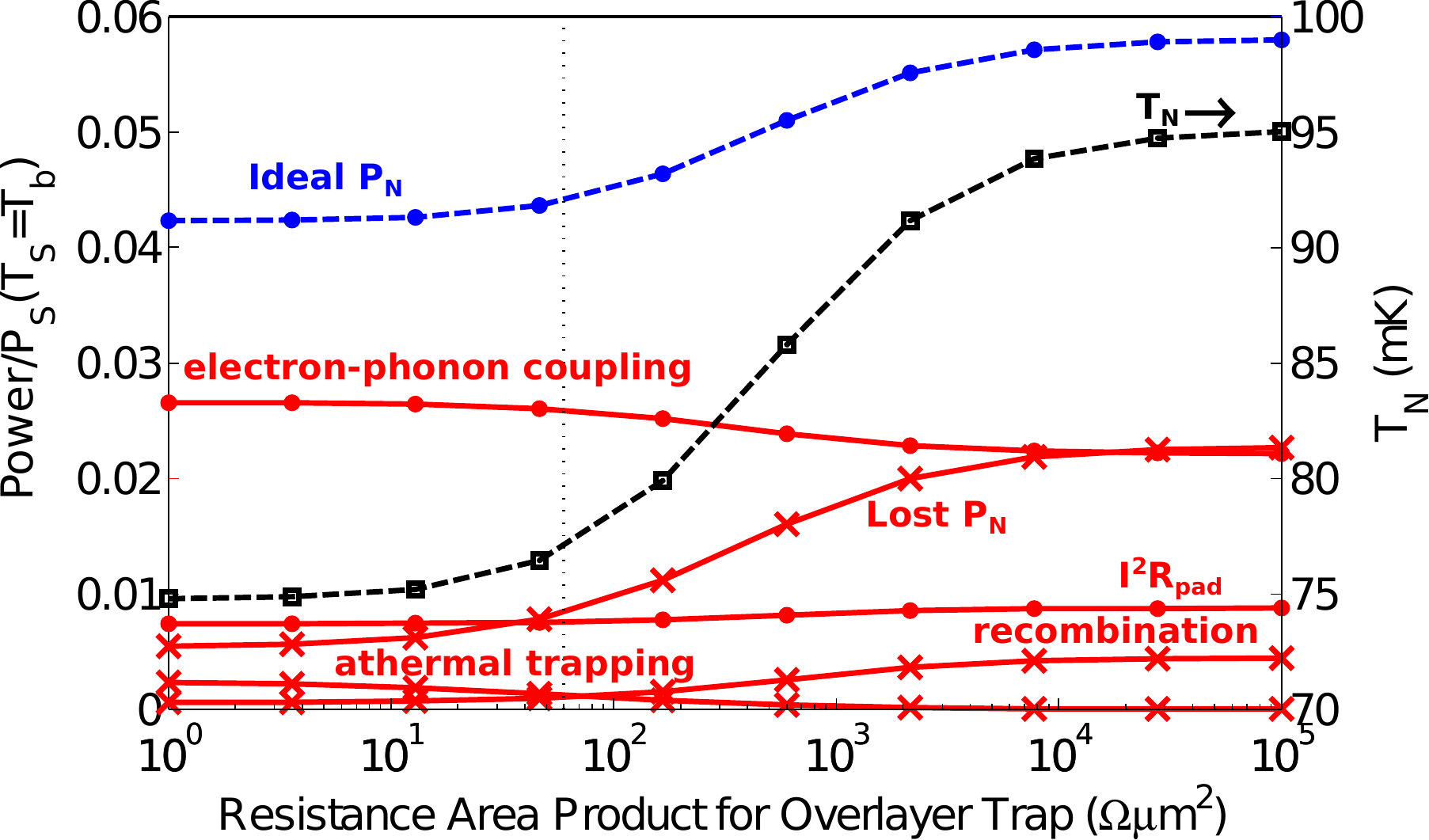}
\caption{\label{fig:300RSP_overlayertrap} Model sweep over the value of the resistance area product for the overlayer trap barrier.  Large resistance area product values represent the case of no overlayer trap, while low values represent a highly transparent trap barrier.  The right vertical axis shows values of $T_N$, the left vertical shows the power loads from Eq \ref{eq_powerbalance_beta} as equivalent $\beta$ values.  The solid red terms sum to equal `Ideal $P_N$'. `Ideal $P_N$' is $P_N|_{T_S=T_b}$ and `Lost $P_N$' is the difference between `Ideal $P_N$' and the actual $P_N$ due to heating of the $S$ layer. The terms with \textcolor{red}{$\times$} markers are often represented as $\beta P_S$.  Decreasing the trap resistance area product causes the $S$ layer temperature $T_S$ to remain closer to the bath temperature $T_B$, reducing `Lost $P_N$' and recombination, and resulting in lower base temperature $T_N$.  Changes in $T_N$ lead to changes in `Ideal $P_N$', the electron-phonon coupling, and the power load due to the athermal behavior of trapped quasiparticles. The majority of the benefit of the overlayer trap is achieved with a resistance area product of $10-100~\Omega \mu m^2$, where $T_N$ values near 75~mK are reached from $T_b=300$~mK.  `Lost $P_N$' (and therefore $\beta$) remains finite even for low resistance area product overlayer because the overlayer trap heats as well. The Andreev ($I_2 V_b$) power load is not shown; it is roughly constant and never greater than the recombination term.}
\end{figure}

Figure \ref{fig:modeltemps} shows the temperatures of the $S$, $O$ and $P$ layers as calculated by the model for parameters in Table \ref{table:central_params_300mK}.  Figure \ref{fig:300RSP_overlayertrap} shows the terms in the normal metal power balance (Eq \ref{eq_powerbalance_beta}) calculated by the model for different values of the overlayer trap barrier resistance area product $\mathcal{R_{SO}}$.

\section{Superconductors and tunneling}
\label{sec:superconductors_and_tunneling}

\subsection{Quasiparticle density and density of states}

This section will briefly review a subset of properties of superconductors which are vital for understanding their behavior in NIS refrigerators.  In a superconductor at zero temperature, all the conduction electrons form Cooper pairs which can carry electrical current with zero resistance.  At finite temperature, some of the Cooper pairs are broken and each broken Cooper pair yields two quasiparticles.  The effective density of states of quasiparticles is given by $N(0)\nu(E)$, where
\begin{equation}
\label{eq:dynes_dos}
\nu(E) = \left| \mathrm{Re}\left(\frac{E/\Delta-i\gamma}{\sqrt{ (E/\Delta-i\gamma)^2-1}}\right) \right|,
\end{equation}
$N(0)$ is the two-spin density of states at the Fermi energy in the same material in the normal state,  $E$ is the energy relative to the Fermi energy, $\Delta$ is the BCS energy gap, and $\gamma$ is a unitless factor that describe deviations from the ideal BCS superconducting density of states.  The parameter $\gamma$ introduces states below the gap energy $\Delta$ and is used to account for observed sub-gap tunneling currents greater those predicted with the ideal BCS density of states ($\gamma=0$).  Pekola et al\cite{PhysRevLett.105.026803}  suggest environment-assisted tunneling as the mechanism responsible for finite $\gamma$.

The quasiparticle density $n$ in a superconductor at temperature $T_x$ is 
\begin{equation}
n(T_x) = 2 N(0)\int_0^{\infty} f_x(E) \nu(E)|_{\gamma=0} dE
\label{eq_qp_density}
\end{equation}
where $f_x(E)=(e^{E/k_b T_x}+1)^{-1}$ is the Fermi function at temperature $T_x$,  and $k_b$ is Boltzmann's constant.  To account for equilibrium behavior, we write the total quasiparticle density as the sum of the thermal density $n_{th}$ at the cryostat bath temperature and the excess density $n_{ex}$.  Relevant quasiparticle distributions are strongly peaked at energy $\Delta$, thus for the purpose of tracking energy during quasiparticle relaxation we treat all quasiparticles as having energy $\Delta$. When we describe the superconductor as having a temperature $T_S\neq T_b$, we choose $T_S$ to have the correct quasiparticle density.

\subsection{NIS Tunneling}

The current in an NIS junction is made up of two parts
\begin{align}
\label{eq:iv_i}
I_{NS} =& I_1 + I_2 , \\
I_1 =&  \frac{1}{q_e R_{NS}} \int_{0}^{\infty}\nu(E)[f_N(E-q_eV_b)\\&-f_N(E+q_eV_b)] dE, \notag\\
\label{eq:twoparticletunneling}
I_2 =& I_{2|N} + I_{2|S},\\ 
\label{eq:i2normal}
I_{2|N} =& \frac{1}{q_e R_{NS}} \frac{\hbar}{q_e^2 \mathcal{R}_{NS} t_N }\tanh{\frac{q_e V_b}{2 k_b T_N}},
\end{align}
where $I_1$ is the single particle tunneling current, $I_2$ is the two particle tunneling current, $I_{2|N}$ is the two particle contribution from the normal metal electrode in the simplest geometry of an infinite uniform junction,\cite{Hekking1994} $I_{2|S}$ is the two particle contribution from the superconducting electrode, $q_e$ is the electron charge, $R_{NS}$ is the tunneling resistance, $\mathcal{R}_{NS}$ is the product of the tunneling resistance and junction area, and $V_b$ is the voltage difference across the junction.  

The form and magnitude of the two particle tunneling current depends on electron interference due to multiple scatterings, and therefore, on the geometry of the electrodes.  We rely on theoretical forms of $I_{2|N}$ and $I_{2|S}$ from Hekking and Nazarov\cite{Hekking1994} which are not specific to our junction geometries.  The form of $I_2$ in Eqs \ref{eq:twoparticletunneling} and \ref{eq:i2normal} is in rough agreement with our data, and we have an ongoing investigation to improve our understanding of the precise form of $I_2$ for our geometry.\cite{Lowell2011}  We have excluded the contribution $I_{2|S}$ from the superconducting electrode in this work because 1) $I_{2|S}\ll I_{2|N}$ due to the thickness of the superconductor 2) the overlayer traps should further suppress multiple reflections and thus the magnitude of $I_{2|S}$ and 3) the theory of $I_{2|S}$ in Hekking and Nazarov\cite{Hekking1994} breaks down for biases $q_e V_b \approx \Delta$, which are commonly used for cooling.   In most cases $I_1 \gg I_2$, however at low temperatures and low biases $I_2 > I_1$.  For the junctions described in this work $I_2$ plays a small role; $I_2$ is more important for junctions with lower resistance area products.

The complete current-voltage (IV) relationship also includes a resistive voltage due to the normal metal electrode
\begin{equation}
\label{eq:iv_v}
V = V_b + I_{NS} R_{pad},
\end{equation}

The power deposited in the normal metal by single particle tunneling is
\begin{align}
\label{eq_pn}
P_N =& \frac{1}{q_e^2 R_{NS}} \int_{-\infty}^{\infty}(q_e V_b-E)\nu(E)[f_N(E-q_eV_b)\\&-f_S(E)] dE.\notag
\end{align}
Refrigeration is possible because $P_N$ is negative for biases such that $q_e V_b \lessapprox \Delta$.  Note that $P_N$ is a function of both $T_N$ and $T_S$ (through $f_N$ and $f_S$), and this dependence on $T_S$ is the reason that superconductor heating directly impacts NIS performance.  The two particle tunneling current deposits power $I_2 V_b$ and Joule heating deposits power $I_{NS}^2 R_{pad}$ in the normal metal.\cite{PhysRevLett.100.207002}

The power deposited in the superconductor is 
\begin{equation}
\label{eq:ps}
P_S = I_{1}V_b-P_N.
\end{equation}
Both the quasiparticle thermal population, and the quasiparticles injected into the $S$ layer by the NIS junction are strongly peaked at $\Delta$.  We approximate the quasiparticle injection rate as $P_S\Delta^{-1}$, which is justified because $T_N\ll\Delta/k_b$ and $eV_b<\Delta$ in the regime of interest for NIS refrigerators.  The power deposited in the superconductor per unit area $\mathcal{P}_S$ is calculated by substituting the resistance area product $\mathcal{R}_{NS}$ of the tunnel junction in place of the resistance $R_{NS}$.

\section{Quasiparticle trapping}
\label{sec:qp_trapping}
\subsection{NIS junction as a quasiparticle trap}

Power will flow across an NIS junction even when $V_b=0$, if the normal metal temperature $T_N$ and superconductor temperature $T_S$ are unequal.  In NIS refrigerator operation, the $S$ layer is directly heated and therefore is hotter than the $O$ layer.  As a result, power will flow from the $S$ to the $O$ layer, providing an additional mechanism for the superconductor to reach thermal equilibrium.  The unbiased NIS junction between the $S$ and $O$ layers, combined with the $O$ layer form a quasiparticle trap used to heatsink the superconductor.  The power flow per unit area from the superconductor is given by
\begin{align}
\label{eq_ptrap_full}
\mathcal{P}_{trap} &= \frac{2}{q_e^2 \mathcal{R}_{trap}} \int_{0}^{\infty}E\nu(E)[f_S(E)-f_N(E)] dE\\
\label{eq_ptrap}
&\approx \frac{\Delta}{q_e^2 \mathcal{R}_{trap} N(0)} [n(T_S)-n(T_O)]
\end{align}
where we have used the approximation that all quasiparticles have energy $\Delta$ to obtain Eq \ref{eq_ptrap}.

The tunneling lifetime of an electron in a thin metal film adjacent to a tunnel barrier is
\begin{align}
\label{eq_tau_tun}
\tau_{t}&=N(0) q_e^2 t \mathcal{R}_{t}
\end{align}
where $t$ is the film thickness and $\mathcal{R}_{t}$ is the resistance area product of the tunneling barrier.  The tunneling lifetime grows with film thickness and also with the resistance area product of the tunnel junction.   The lifetime for quasiparticles to trap from the $S$ to the $O$, or to tunnel back from the $O$ to the $S$ layer is $\tau_t=104$~ns, based upon thickness $t\approx500$~nm, and a tunnel barrier with resistance area product $\mathcal{R}_{t}=60~\Omega \mu m^2$.

\subsection{What happens to a trapped quasiparticle?}
\label{section_what_happens_trapped_qp}
A quasiparticle which tunnels from the $S$ into the $O$ layer becomes an excited electron with energy $\Delta$.  There are three processes available to this electron: 1) tunneling back into the superconductor with lifetime $\tau_{tun}=104$~ns, 2) scattering with another electron with lifetime $\tau_{e-e|\Delta}=540~$ns, and 3) scatter and create a phonon with lifetime $\tau_{e-p|\Delta}=25$~ns.  Once the excited electron has scattered with either a phonon or an electron, it will have energy less than $\Delta$ and be unable to tunnel back into the superconductor, thus the name quasiparticle trap.\cite{Booth1987} However, the most likely phonon to be created will have energy $3\Delta/4$ which is well above the thermal distribution.\cite{Ullom1999}  A $3\Delta/4$ phonon has a finite probability of being absorbed in the $N$ layer, and if this happens regularly it will severely degrade the benefit of quasiparticle traps.  Appendices \ref{section:diffusion}--\ref{sec:eescattering} describe these mechanisms and the methods used to calculate the time constants.

We estimate the probability of scattering to create a phonon $\Pi_p$, to tunnel back to the superconductor $\Pi_{tun}$, or to undergo electron-electron scattering $\Pi_e$ as
\begin{align}
\label{eq:pi_all}
\Pi_{(p)/(tun)/(e)} &= \frac{\tau_{(e-p|\Delta)/(tun)/(e-e|\Delta)}^{-1}}{\tau_{e-p|\Delta}^{-1}+\tau_{tun}^{-1}+\tau_{e-e|\Delta}^{-1}}.
\end{align}
For $\Delta=190~\mu$eV, $\Pi_p=71\%$ of the quasiparticles removed from the $S$ layer will generate athermal phonons, $\Pi_{tun}=24\%$ tunnel back to the superconductor, and only $\Pi_e=5\%$ are thermalized by electron-electron scattering. The most likely energy for these athermal phonons is $E=3\Delta/4$, so $\sim\frac{3}{4}\Pi_p=53\%$ of the energy which tunnels from $S$ to $O$ becomes athermal phonons.    Appendix \ref{section_athermalphonons} describes a ray tracing model which predicts the fraction of these athermal phonons that deposit their energy in the other layers.  The majority of athermal phonons will be reabsorbed before leaving the $O$ layer, making the overlayer traps more effective than these probabilities suggest on their own.

\section{Experimental Methods and Results}
\label{sec:experiment}
SINIS refrigerator devices were fabricated on a Si wafer with $\sim$150~nm of thermally grown SiO$_2$.  The $N$ layer is $\sim$20~nm of sputter deposited Al with $\sim4000$~ppma\footnote{Parts per million by atomic percent (ppma) is used in this work, however parts per million by weight (ppmw) is commonly used by commercial vendors and analysis labs.} Mn.  The $N$ layer and all subsequent layers are patterned with standard photolithography techniques.  The metal layers are etched with an acid etch.  A layer of SiO$_2$ is deposited by plasma enhanced chemical vapor deposition and etched with a plasma etch.  The vias etched in the SiO$_2$ define the location of the tunnel junctions.  The wafer is returned to the deposition system where it is ion milled to remove the native oxide from the AlMn, then exposed to a linearly increasing pressure of 99.999$\%$ pure O$_2$ gas for 510 seconds, reaching a maximum pressure of  6.0~Torr (800~Pa) to form the tunnel barrier oxide.  Then, $\sim$500~nm of Al is deposited by sputter deposition to form the $S$ layer.  The overlayer trap tunnel barrier is formed by exposure to 32~mTorr O$_2$ (4.3~Pa) for 30~s, and $\sim$500~nm of AlMn is deposited to form the $O$ layer.

\begin{figure}
\includegraphics[width=0.9\linewidth]{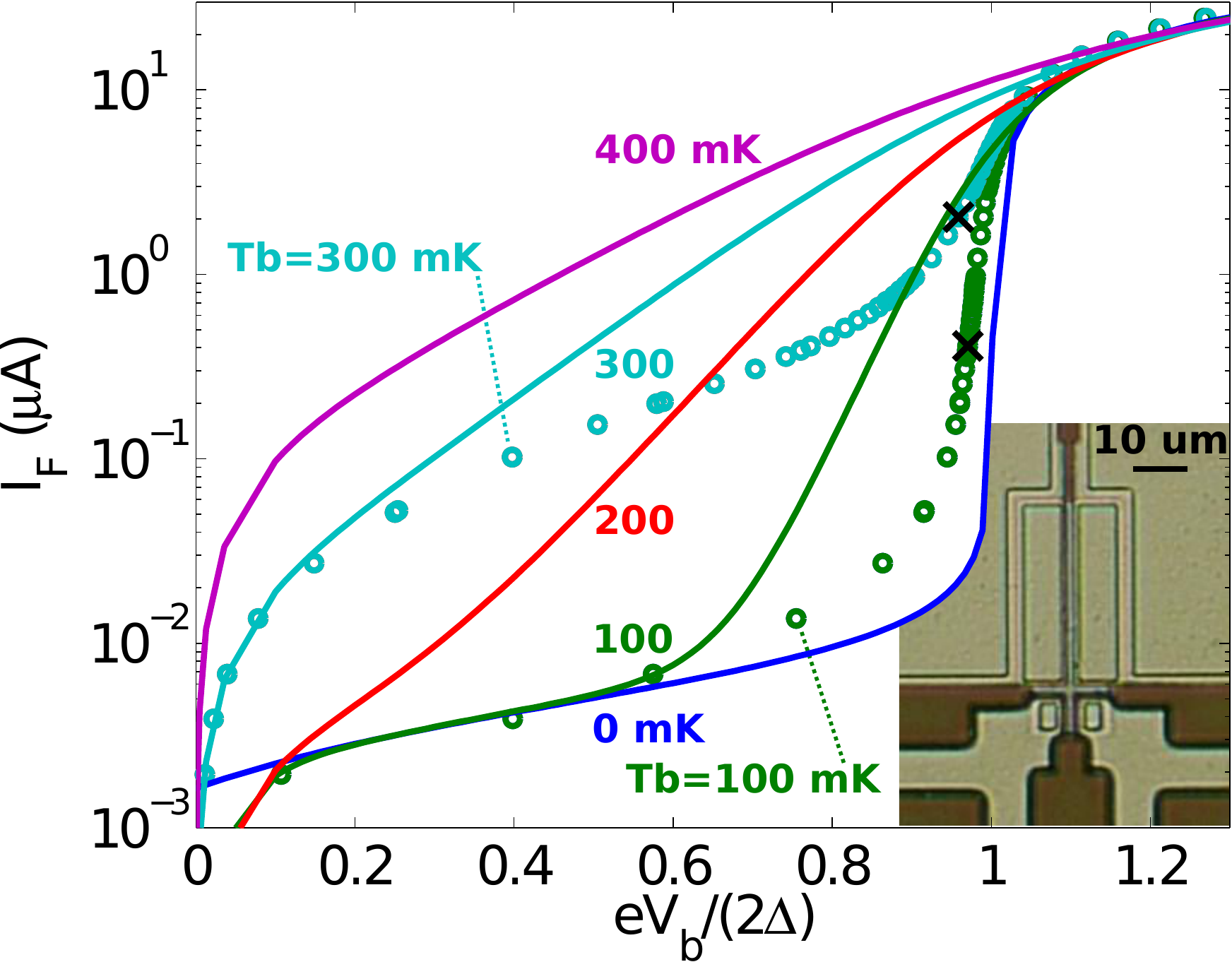}
\caption{\label{fig:ivcal} Theoretical isothermal NIS current-voltage curves from Eqs \ref{eq:iv_i} and \ref{eq:iv_v} shown as solid lines.  Measured NIS refrigerator IV curves taken at 100~mK and 300~mK bath temperatures  shown as circles.  For each data point, the temperature $T_N$ is uniquely determined by the theory curve upon which it falls.  The optimal bias in the data for each bath temperature is shown as an $\times$ marker.  The inset shows a picture of the device.  The two large junctions (vertical rectangles) are the refrigerator junctions. The two smaller junctions are independent thermometer junctions.}
\end{figure}

\begin{figure}
\includegraphics[width=0.9\linewidth]{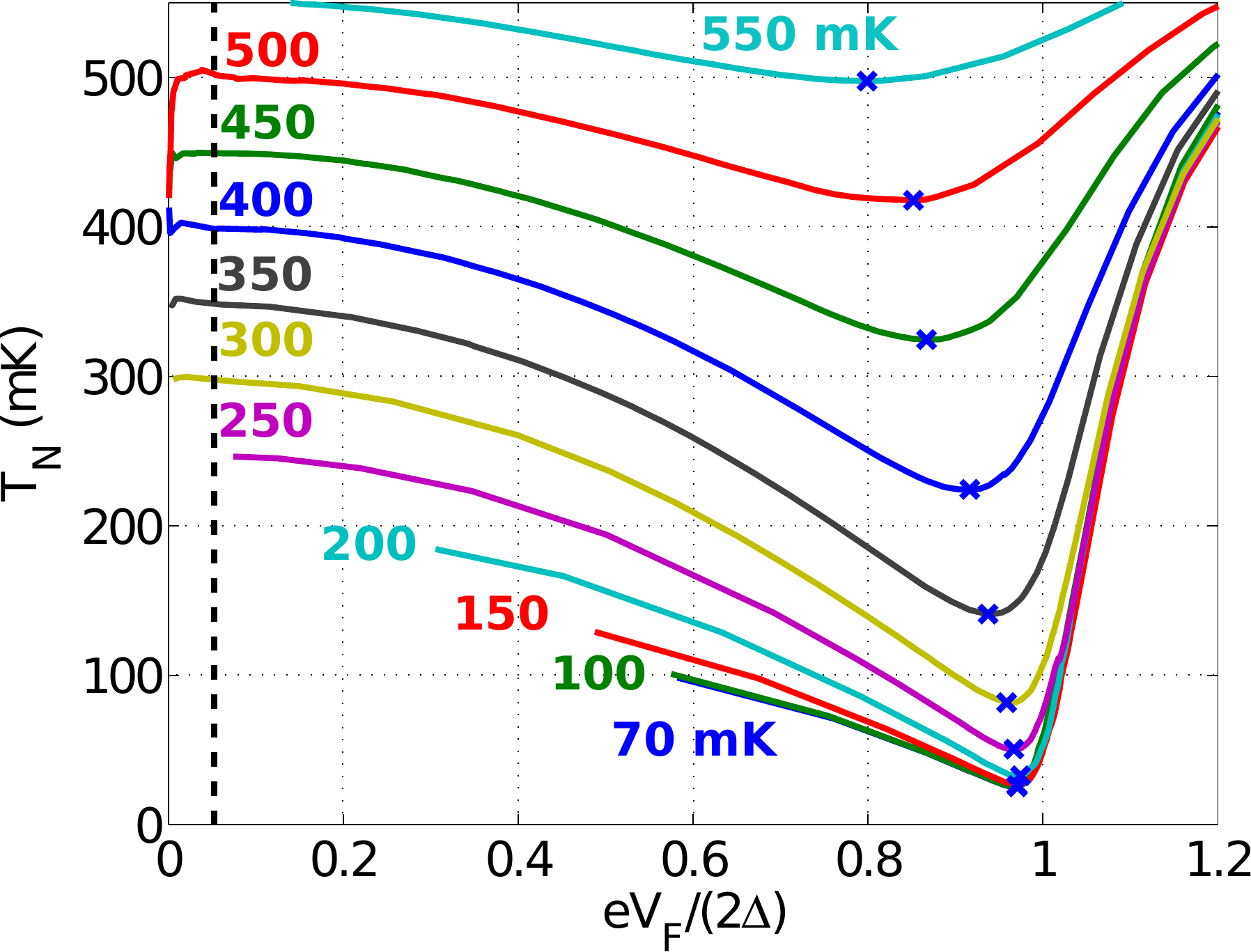}
\caption{\label{fig:tempvsbias} Temperature $T_N$ determined from the refrigerator junctions vs NIS refrigerator voltage bias as solid lines for many bath temperatures $T_b$.  The vertical dashed line at 20~$\mu$V is used to determine uncooled NIS data points, for use in calibrating the conversion between $V_F$ and $T_N$.  The optimal bias points are shown as \textcolor{blue}{$\times$}, and indicate lowest temperature $T_N$ achieved at each bath temperature $T_b$.    }
\end{figure}

One of these devices (inset in Fig \ref{fig:ivcal}) was cooled in a copper box attached to the cold stage of an adiabatic demagnetization refrigerator. We measured current-voltage ($IV$) curves of the SINIS device (and the independent thermometer junctions) at many bath temperatures $T_b$. The bias of the refrigerators (independent thermometers) was set with a computer-controlled, battery-powered voltage source connected through a 100~k$\Omega$ (10~M$\Omega$) bias resistor.  The voltage across the device was amplified with an operational amplifier and the output of the amplifier was measured by a digital multimeter.  The current was calculated as the source voltage minus the device voltage divided by the bias resistance. 

The primary experimental results we report come from thermometry based on the refrigerator junctions.  Each data point ($I$ and $V$) is an independent measurement of the normal metal temperature $T_N$, which depends on both bath temperature $T_b$ and refrigerator bias $V$.  The temperature $T_N$ at each point is uniquely determined by comparison to isothermal theory curves (Eqs \ref{eq:iv_i} and \ref{eq:iv_v}), as shown in Fig \ref{fig:ivcal}.  Temperatures determined from NIS junctions in this way are effective temperatures because Fermi occupation functions are assumed in the junction electrodes. For a thermal electron distribution this effective temperature will match the actual temperature. For athermal electron distributions, the NIS effective temperature does not provide a complete description.   Figure \ref{fig:tempvsbias} shows effective temperature $T_N$ vs refrigerator bias for many bath temperatures.

\begin{figure}
\includegraphics[width=0.9\linewidth]{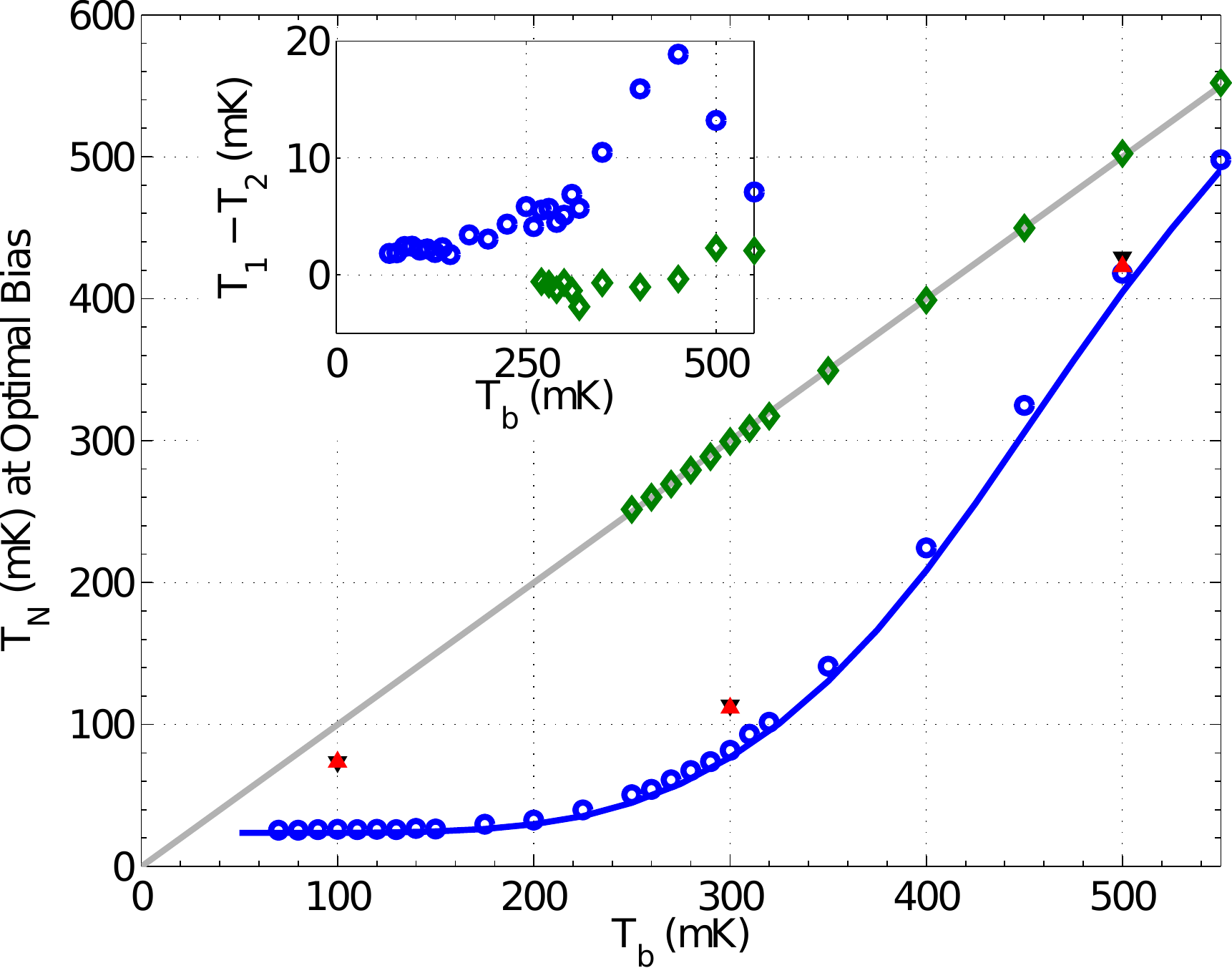}
\caption{\label{fig:temps} Temperature of the normal metal $T_N$ determined from refrigerator junctions at optimal bias voltage   as \textcolor{blue}{$\circ$} vs bath temperature $T_b$.  Temperature $T_N$ from the independent thermometer junctions as $\blacktriangledown$, and the temperature at the refrigerator junction calculated from thermometer junction temperature as \textcolor{red}{$\blacktriangle$} (see Sec \ref{sec:athermal}).  Uncooled temperature $T_N$ at $V_F=20~\mu$V as \textcolor{green}{$\Diamond$}.  The solid blue line is the predicted temperature $T_N$ at optimal bias and the solid grey line is $T_N=T_b$, which represents zero heating or cooling.  The inset shows the difference between the model and measured $T_N$ at optimal bias as \textcolor{blue}{$\circ$}, and a subset of the difference between the temperature $T_N$ at $V_F=20~\mu$V and $T_b$ as \textcolor{green}{$\Diamond$}.  Uncooled temperature points at bath temperature below 250~mK are excluded because they are extremely sensitive to the form of the two particle tunneling current, which is not fully understood.  The theory and data for $T_N$ at optimal bias are in very good agreement.  These results are an improvement over previous cooling with large area NIS junction refrigerators.\cite{Clark2004}    }
\end{figure}

For each bath temperature, we determine the bias point at which the minimum value of $T_N$ is reached.  We call this the optimal bias.  The value of $T_N$ at the optimal bias is shown vs bath temperature in Fig \ref{fig:temps}.  Figure \ref{fig:temps} also shows the value of $T_N$ at a bias voltage of 20~$\mu V$.  For bath temperatures above $\sim250$~mK the temperature $T_N$ determined at a bias voltage of 20~$\mu V$ should equal the bath temperature because the NIS junctions have little thermal effect at low biases.  We determine $\Delta$ by a least squares minimization between the temperature deduced from IV curves at low bias (20~$\mu V$) and the temperature measured by the cryostat thermometer.  The $N$ layer resistance $R_{pad}$ is calculated as in Table \ref{table:central_params_300mK}, the tunneling resistance $R_T$ is the asymptotic resistance at high bias and high temperature minus $R_{pad}$, and $\gamma$ is chosen to match the maximum differential resistance at 70~mK.  

The data from the refrigerator IVs indicate significantly improved cooling compared to previous large area ($>100~\mu$m$^2$) NIS refrigerators.\cite{Clark2004} Assuming a thermal electron distribution in the normal metal, we deduce cooling from 300~mK to 82~mK and from 100~mK to 26~mK.

\section{Independent thermometers and athermal electron distributions}
\label{sec:athermal}

Current-voltage curves from the independent thermometers can be used to obtain temperature values using a procedure similar to that described in the previous section.  As shown in Fig 7, the thermometer data suggests hotter electron temperatures than those deduced from the refrigerators.   The thermometer junctions were fabricated using a double oxidation technique and their resistance area product was 145,000~$\Omega \mu$m$^2$, over 100 times greater than the refrigerators.\cite{Holmqvist2008,O'Neil2011}  This high resistance makes the thermometers thermally neutral, meaning they neither heat nor cool the normal metal.  We next consider thermal mechanisms for a temperature gradient between the thermometers and refrigerators. 

We expect a small temperature difference between the refrigerator and thermometer junctions due to the finite thermal conductivity of the normal electrode and the presence of power loads within it. We calculate the expected difference by solving the heat equation in the N layer including electron-phonon coupling and stray power:
\begin{equation}
\kappa_N \frac{d^2 T_N}{dx^2} = \Sigma(T_N^6-T_b^6) - P_{excess}/U_N
\end{equation}
We let x vary from 0 to 10 microns with $x=0$ corresponding to the edge of the refrigerator junction, $x=10$ corresponding to the end of the AlMn, and $x=4$ to $9$ corresponding to the thermometer junction.  We fix $dT/dx$ at $x=0$ at $x=10$ and vary the refrigerator temperature at $x=0$ until the thermometer temperature (evaluated at $x=8$) matches the measured temperature. The variable $\kappa_N$ is the thermal conductivity of the normal layer and is obtained from Eq \ref{eq_weidemannfranz}.  The stray power $P_{excess}=0.05$~pW is obtained from thermometer junction temperature measurements with the refrigerator junctions at zero bias.  Finally, $U_N$ is the N layer volume. The observed thermometer temperatures and the refrigerator temperatures calculated from these thermometer temperatures are shown in Fig \ref{fig:temps}. The gradient varies from ~0.02 mK to ~7 mK for bath temperatures between 100 mK and 500 mK. This gradient accounts for over half of the observed difference at a bath temperature of 500 mK but does not explain the observed difference near 100 mK. We hypothesize that the remaining differences in temperature result from an athermal electron distribution created by the refrigerators that recovers to a thermal distribution over the distance to the thermometers.

NIS refrigerators generate an athermal electron distribution when the tunneling rate is faster than inelastic scattering in the normal electrode.   The tunneling time for electrons above the gap edge is insensitive to temperature, and the rates for inelastic electron-electron and electron-phonon scattering both decrease with temperature.  As a result, athermal effects in NIS junctions appear near or below $\sim100$~mK.\cite{Pekola2004}  The athermal electron distribution created by the refrigerators lacks the high energy excitations ordinarily found in the tips of a Fermi distribution.  However, its total excitation energy will exceed that of the thermal distribution that results in the same tunneling current. 

Over the distance between the refrigerator and thermometer junctions, the athermal distribution recovers to a thermal one through inelastic scattering and the tips of the Fermi distribution are repopulated. Consequently, additional current flows through the thermometers and a higher device temperature is deduced even though the thermal electron distribution at the thermometers and the athermal distribution at the refrigerators contain the same excitation energy (neglecting the small gradients calculated above).  We observe that the temperature deduced from the thermometers is largely independent of their bias point, supporting the notion of a thermal electron distribution at the thermometer junctions.  Further, athermal electron distributions in metals with similar diffusion constants to our normal layer have been observed to thermalize over distances as short as 2.5~$\mu$m.\cite{PhysRevLett.79.3490}

Our results illustrate the complexities of junction thermometry.  Experiments that use the same junctions as both refrigerators and thermometers are susceptible to athermal effects.  Similarly, if independent thermometer junctions are close enough to the refrigerators that they sample the same athermal distribution, the inferred temperature reduction may be exaggerated.  We have provided temperature values deduced from both refrigerator and distant thermometer junctions.  While the thermometers likely provide a truer measurement of temperature, the results from the refrigerators still have value as a measure of the distortion of the electron distribution in our devices. 

We note that our equilibrium thermal model predicts temperatures close to the values deduced from the refrigerator IVs, rather than the independent thermometers.  This outcome is not surprising.  While the model finds the temperature where the power loads balance, it can also be thought of as finding the junction current at thermal balance.  In the presence of an athermal electronic distribution, current remains a good predictor of key power loads such as the Joule term and the heating of the superconductor. Hence, the model correctly predicts refrigerator currents together with the temperature that produces these currents assuming a thermal distribution.  Since the refrigerator temperatures are deduced from measured currents assuming a thermal distribution, model and refrigerator measurements give similar temperature values.

\section{Analysis and comparison to model}

\begin{figure}
\includegraphics[width=0.9\linewidth]{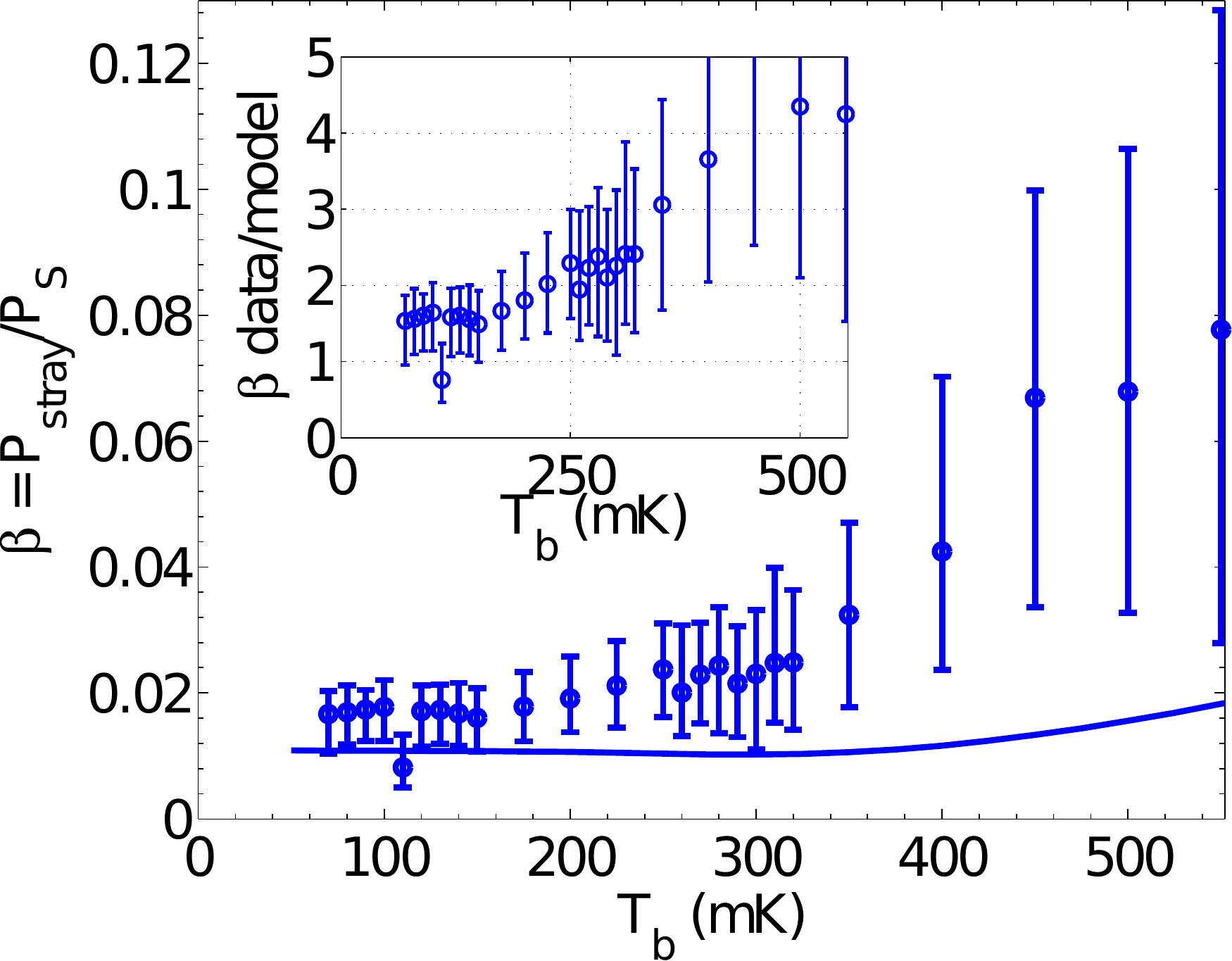}
\caption{\label{fig:beta} Measured values of $\beta$ vs $T_b$ as circles and model predictions as solid line.  The inset shows the ratio between the measured and predicted value vs $T_b$.  Previous predictions of $\beta$ or equivalent parameters have varied from measured values by factors on order $\sim3-10$, so a factor of 1.5 is considered good.  The model predicts the cooling to within a few mK at low temperatures, which is useful for designing improved NIS refrigerators. }
\end{figure}

Since knowledge of the precise details of the electron distribution under refrigeration is lacking, for the purposes of this section we henceforth assume a Fermi distribution in the normal metal and continue our analysis of the results from that perspective. We calculate $\beta$ vs bath temperature and compare to predictions of the thermal model, results are shown in Fig \ref{fig:beta}. In order to determine the value of $\beta$ in the device, we calculate the excess power load required to explain the measured value of $T_N$ at the optimal bias at each bath temperature;  
\begin{equation}
\begin{split}
P_{excess} = &-2 P_N(R_T/2,\Delta,V_o/2, T_N, T_b) \\&- I_o^2 R_{pad} - U_N \Sigma (T_b^6-T_N^6)-I_2 V_o
\end{split}
\end{equation}
where $R_T$ is the tunneling resistance of the two refrigerator junctions in series, the factors of 2 account for the existence of two junctions, $P_N$ is given in Eq \ref{eq_pn}, $V_o$ and $I_o$ are the optimal bias voltage and current for a given bath temperature, $T_N$ is the $N$ layer electron temperature at the optimal bias, $R_{pad}$ is the resistance of the $N$ layer current path, $U_N$ is the $N$ layer volume, and $\Sigma$ is the electron phonon coupling constant.  The value of $\beta$ is calculated by
\begin{equation}
\beta = \frac{P_{excess}}{I_o V_o - 2 P_N(R_T/2,\Delta,V_o/2, T_N, T_b)},
\end{equation}
where $I_o V_o$ is the dissipated power, and the denominator is the total power deposited in the $S$ layer.  The measured value of $\beta$ is therefore a complicated combination of measurement and analysis that depends on the parameters $\Delta$, $\Sigma$, $\gamma$, $U_N$, $R_T$ and $R_{pad}$, and the accuracy of the underlying theory.  We generate error bars on $\beta$ using estimates of the uncertainty of each parameter propagated through the calculation. The uncertainties used to generate the error bars  are $\Delta:\pm0.25\%, \Sigma:\pm20\%, \gamma:\pm10\%, U_N:\pm5\%, $and $R_{pad}:\pm10\%$.  The sum of $R_{pad}$ and $R_T$ was held constant.  Variation in $\Delta$ affects the determination of $T_N$ and directly enters the calculations of $P_N$.  At higher temperatures uncertainty in $\Sigma$ determines the bulk of the error bars; at lower temperatures uncertainty in $\Delta$ dominates. The range of 300~mK and below is the most interesting range for Al-based NIS refrigerators.  In this range, the measured value of $\beta$ is roughly 1.5 times the predicted value, which is better agreement than past NIS modeling has achieved. For higher temperatures, the agreement is not as good, suggesting that improvement could be made by focusing on features of the model whose relative importance increases with temperature, such as quasiparticle recombination and electron-phonon coupling.  Underwood et al\cite{Underwood2011} observe that the exponent in the electron phonon coupling depends on temperature for temperatures above $\sim300$~mK; including this effect may improve agreement at higher temperatures.

\section{Future Refrigerator designs based on thermal model}

By exploring the design parameter space with the thermal model, we have developed an NIS design that is predicted to cool from 100~mK to 6.5~mK.  The overlayer trap resistance $\mathcal{R}_O$ is decreased to 1~$\Omega \mu m^2$, the overlayer thickness $t_O$ is increased to 10~$\mu m$, the Dynes parameter $\gamma$ is reduced to the lowest reported value\cite{PhysRevLett.105.026803} of 38000$^{-1}$, and the junction length parallel to current flow $x_j$ is reduced to 3~$\mu m$.  

The same changes that we suggest to improve cooling from 100~mK will also improve cooling from higher temperatures with lower resistance area product refrigerator junctions.  A device with the changes above, increased $N$ layer thickness of $t_N=30$~nm, and decreased the resistance area product of $\mathcal{R}_{NS}$=150~$\Omega \mu m^2$ is predicted to cool from 300~mK to 48~mK.  If the overlayer thickness is left unchanged at 500~nm, the device should cool to 64~mK.  Junctions with this low resistance area product are difficult to fabricate with high yield, so it is not a simple step to achieve these performance levels. These predictions demonstrate the capability of the model to explore a large design space. Refrigerators based on these designs will generate electron distributions that vary further from the Fermi distribution than the refrigerator discussed here.

\section{Conclusions}

We have described a comprehensive NIS refrigerator thermal model that includes differential equations for three temperatures vs position, and accounts for the athermal behavior of high energy excitations.  The predictions of this zero free parameter model agree well with experimental results from an NIS refrigerator with overlayer traps.  The agreement for the quasiparticle return parameter $\beta$ are within a factor of 1.5 over the temperature range of primary interest. This refrigerator cools more than previous large area ($>100\mu$m$^2$) NIS refrigerators.  From a bath temperature of 300~mK, the refrigerator junction current is consistent with cooling to 82~mK. However, the refrigerators likely generate an athermal electron distribution, and as a result the cooling cannot be described simply.  Thermometer junctions which probe the temperature $\sim8$~$\mu$m away from the athermal distribution measure a temperature 114~mK. 

This work expands upon previous modeling efforts by including more coupled temperatures systems (the overlayer electrons and the metal layer phonons) and more athermal effects. The model could be further improved by including more detail in the form of the electron-phonon coupling, and by calculating the electron distribution in the normal metal electrode. We plan to perform phonon cooling experiments that will provide a direct measurement of the usable temperature reduction from similar refrigerator junctions.

\acknowledgements{We benefited from the help of many other members of the Quantum Sensors Program and Quantum Fabrication Facility.  This work was funded by the NASA APRA Program and the NRC RAP.}

\appendix

\section{Electron excitation behavior in a normal metal}
\label{appendix1}
To calculate the temperature of the $S$ layer, the $O$ layer, and the effect of athermal excitations in both layers, we must understand the heat transport and loss mechanisms in both a superconductor and a normal metal.  In particular, we want to understand what happens when a quasiparticle is trapped in the $O$ layer.

\subsection{Electron diffusion in the normal metal}
\label{section:diffusion}
The numerous elastic collisions off impurities, grain boundaries, and surfaces cause electrons to diffuse, rather than move ballistically, through a normal metal.  The diffusion constant for electrons in a normal metal is\cite{Kittel1996}
\begin{align}
\label{eq_diffusionconstant_normal}
D_x &= \frac{1}{q_e^2 N(0) \rho_x} =\frac{l_x v_F}{3}
\end{align}
where $l_x$ is the elastic mean free path and $v_F=2.02\times10^6$~m/s is the Fermi velocity for Al.  From this we calculate the elastic mean free paths in the layers of the NIS refrigerator to be $l_N=6$~nm, $l_O=9$~nm and $l_{S}=340$~nm.

\subsection{Thermal transport in normal metal electrons}
\label{section:thermal_transport}
The Wiedemann-Franz law relates the thermal conductivity of the electron system to the resistivity by
\begin{equation}
\label{eq_weidemannfranz}
\kappa_x = \frac{L T_x}{\rho_x}
\end{equation}
where  $L=2.45\times 10^{-8}~$W $\Omega$/K$^2$ is the Lorenz number.\cite{Kittel1996}  


\subsection{Electron-phonon interaction in normal metals}
\label{section_ep_epcoupling}
Electron and phonon populations occupying the same volume of a normal metal interact via the electric field from phonons displacing lattice ions.\cite{Reizer1989,Ullom1999}  This interaction leads to a power per unit volume flowing from the phonon system to the electron system of the form
\begin{equation}
\label{ep_epcoupling}
\mathcal{P}_{p-e}=\Sigma (T_P^n-T_e^n),
\end{equation}
where $T_P$ is the phonon temperature, $T_e$ is the electron temperature, $\Sigma$ is the electron-phonon coupling constant, and $n$ is an exponent whose value can vary from 4 to 6 depending on the properties of the metal film.  Previous work on NIS junctions with AlMn base electrodes has used $n=5$.\cite{Clark2005} However, recent measurements by Underwood\cite{Underwood2011} and Taskinen\cite{Taskinen2006} show that $n=6$ is more accurate for AlMn films in the temperature range of interest.

 
\subsubsection{Electron-phonon interaction: Electron scattering time $\tau_{e-p}$}
Ullom\cite{Ullom1999} extended theoretical calculations for $P_{p-e}$ to find $\tau_{e-p}$, the characteristic time for an electron to down scatter with the creation a phonon.  In the case where $n$=5 in Eq \ref{ep_epcoupling}
\begin{equation}
\tau_{e-p, n=5} = \frac{3 I_0 k_b^5 N(0)}{\Sigma E^3}
\label{eq_tau_ep_n5}
\end{equation}
where $I_0$=$\Gamma(5)\zeta(5)\approx25$.  

To obtain $\tau_{e-p}$ for the $n=6$ case we use an energy scaling argument to modify Eq \ref{eq_tau_ep_n5}. A method to estimate the electron lifetime for phonon scattering is $\tau_{e-p} \approx C_e/\mathcal{G}_{e-p}$ where $C_e=\gamma T$ is the electron heat capacity\cite{Kittel1996} per unit volume and 
\begin{equation}
\mathcal{G}_{e-p}=d \mathcal{P}_{e-p}/dT_e = n \Sigma T_e^{n-1}
\end{equation}
 is the electron-phonon thermal conductivity.  Thus $\tau_{e-p} \approx \gamma n_d / (n \Sigma T_e^{n-2})$.  This method underestimates $\tau_{e-p}$ because it represents an average of $\tau_{e-p}$ over thermal occupation, and the energy dependence of $\tau_{e-p}$ is strong.  However, this method accurately predicts the energy scaling of $\tau_{e-p}$.  Therefore we modify Eq \ref{eq_tau_ep_n5} for $n=6$ by replacing $\Sigma E^3$ with $\Sigma E^4/k_b$.  The result is
\begin{equation}
\tau_{e-p,n=6} = \frac{3 I_0 k_b^6 N(0)}{\Sigma E^4}
\label{eq_tau_ep}
\end{equation}
We introduce the notation $\tau_{e-p|\Delta}=25$~ns, to represent $\tau_{e-p}$ evaluated at energy $\Delta$. 

An electron scattering with emission of a phonon\cite{Ullom1999} with initial energy $\Delta$ will have average final energy $\Delta/4$ with the emission of a phonon of energy $3\Delta/4$.\footnote{See supplemental material.}  Since this phonon's energy is well above the thermal distribution at 300~mK its athermal behavior must be accounted for, as described in Appendix \ref{section_athermalphonons}.

\subsubsection{Electron-phonon interaction: Phonon scattering time $\tau_{p-e}$}
\label{section:tau_pe}

The energy dependence of the lifetime $\tau_{p-e}$ for a phonon of energy $E$ to lose energy by scattering and promoting an electron is weaker than for $\tau_{e-p}$, so the approximation $\tau_{p-e} \approx C_P/\mathcal{G}_{p-e}$ is more appropriate.  Using the Debye result for phonon heat capacity, 
\begin{equation}
\label{eq_tau_pe_n6}
\tau_{p-e,n=6}=\frac{234 N_A n_d k_b^3}{6 \Sigma \Theta_D^3 E^2}.
\end{equation}

\subsection{Electron-electron scattering}
\label{sec:eescattering}
The electron-electron scattering rate $\tau_{e-e}^{-1}$ is strongly dependent on the effective dimensionality of the sample, and the degree of disorder present.  Therefore, we will first look at the criteria for the 2D/3D limits and the clean/dirty limits.

\subsubsection{2D vs 3D scattering}
The material is in the 2D limit if the excitation energy of the electron is less than the uncertainty in the energy associated with the transit time across the thinnest dimension, that is $\tau_{transit}<\hbar/E$.  Otherwise it is in the 3D limit. All of the films here are thicker than the elastic mean free path so the transit time is $\tau_{transit}=t^2/D$ where $D$ is the diffusion constant. Therefore $\tau_{transit-N}=1.1\times10^{-13}$~s and $\tau_{transit-O}=4.0\times10^{-11}$~s.  The excitation energy of most interest for scattering is $\Delta$, and $\hbar/\Delta=3.5\times10^{-12}$~s.\cite{Ullom1999}  We determine that the $O$ layer is in the 3D limit and the $N$ layer is in the 2D limit for electrons with energy $\Delta$, such as trapped quasiparticles.   


\subsubsection{Electron-electron scattering: clean vs dirty scattering}
 The relevant condition for clean vs dirty electron-electron scattering limits is different in the 2D and 3D cases. In 3D, dirty scattering dominates when $E<E_F/(k_F l)^3$ and in 2D dirty scattering dominates when $E<(\hbar v_F/l)(\pi/(k_F t)) $.  The $N$ layer is in the dirty limit, and the $O$ layer is in the clean limit.
 
\subsubsection{Electron-electron scattering time: $\tau_{e-e}$}
The $O$ layer films look 3D and clean to electrons of energy $\Delta$, and the $N$ layer films look 2D and dirty to the same electrons.  Therefore the electron-electron scattering times are given by
\begin{align}
\label{tau_ee_clean3d}
\tau_{e-e,clean,3D} &= \frac{8 \hbar E_F}{\pi E^2},\\
\label{tau_ee_dirty2d}
\tau_{e-e,dirty,2D} &= \frac{2 \pi \hbar^2 t}{q_e^2 \rho E \ln(k_b T_1/E) }
\end{align}
where $T_1\approx 10^{12}$~K.  When an electron with initial energy $\Delta$ scatters, the average final energy is $\Delta/3$ with the creation of two new excitations, an electron and a hole, each with energy $\Delta/3$.\cite{Ullom1999,Altshuer1979,Santhanam1984}

\section{Excitation behavior in a superconductor}
Having looked at the mechanisms of thermalization that take place in a normal metal, we now examine their analogs in a superconductor. Quasiparticles undergo diffusion by many elastic collisions, much like electrons,  but inelastic scattering is rare for quasiparticles with energy $\sim\Delta$.  Instead, the most common method for quasiparticle relaxation is recombination, whereby two quasiparticles combine to form one Cooper pair and one athermal phonon with energy  $2\Delta$.

\subsection{Quasiparticle diffusion}

We treat all quasiparticles as having equal energy $\Delta$ with one exception.  This exception is the quasiparticle group velocity, which is strongly energy dependent and approaches zero as $E$ approaches $\Delta$.   The group velocity $v_S(E)$ for a quasiparticle excitation of energy $E$ is\cite{Ullom1999,Ullom1998,Bardeen1959}
\begin{align}
v_S(E) &= v_F \sqrt{1-(\Delta/E)^2}\\
\langle v_S(T_S)\rangle&= \frac{2 N(0)}{n(T_S)}\int_0^{\infty} v_S(E)  f_S(E) \nu_0(E).
\end{align}
The diffusion constant for quasiparticles in a superconductor is modified compared to the normal state by this group velocity, such that $D_S=D_{S-normal} \langle v_S\rangle/v_F$.  However, the quasiparticle distribution most relevant to NIS refrigeration is not thermal.

An NIS refrigerator junction injects an athermal distribution of quasiparticles into the $S$ layer.  The diffusion constant for these injected quasiparticles is
\begin{equation}
\begin{split}
\label{eq_qp_diffusion_injection}
D_{S-I} &= \frac{D_{S-normal}}{\Gamma_S(V_b)} \int_{-\infty}^{\infty}\nu(E) \frac{v_S(E)}{v_F} [f_N(E-q_e V_b)\\&+f_N(E-q_e V_b)-2f_S(E)] dE
\end{split}
\end{equation}
where $\Gamma_S$ is the quasiparticle injection rate, which can be calculated with this same integral with $v_S/v_F \rightarrow 1$.  For an NIS refrigerator cooling from 300~mK to 100~mK, $D_{S-normal}/D_{S_I}=20$ and $D_{S-I}=10^{10}~\mu m^2 /s$ are typical values.

\subsection{Quasiparticle-phonon scattering}
Like an excited electron in a normal metal, a quasiparticle can relax to a lower energy state by emitting a phonon.  The lifetime for this is $\tau_{qp-p}$.  High energy $(E>>\Delta)$ quasiparticles relax at a rate almost identical to high energy electrons in a normal metal.  For energies close to $\Delta$,  quasiparticle-phonon scattering is rare.  In Al at a temperature of 250~mK and an excitation energy of $1.2 \Delta$, the scattering time $\tau_{qp-p}$ is many microseconds.  Therefore, $\tau_{qp-p}$ will be much longer than other relevant time constants and quasiparticle-phonon scattering is not considered for the NIS refrigerators discussed here\cite{Ullom1999}.

\subsection{Quasiparticle-quasiparticle recombination}
Two quasiparticles can recombine to form a Cooper pair and emit a phonon with energy equal to the sum of their excitation energies.  This will result in athermal phonons with energy $2\Delta$, whose behavior is described in Appendix \ref{section_athermalphonons}.  Since recombination is a two-body process, the recombination rate scales as the number of pairing possibilities
\begin{equation}
\label{eq_gamma_r}
\frac{dn}{dt} = -\Gamma_R n^2
\end{equation}
where $n$ is the quasiparticle density. The coefficient $\Gamma_R$ depends on material properties, 
\begin{equation}
\Gamma_R=(\frac{\Delta}{k_b T_c})^3 \frac{4}{N_S(0) \Delta \tau_{0-qp}}
\end{equation}
where $\tau_{0-qp}$ is a material specific parameter, whose value we take to be $100$~ns in the $S$ layer.\cite{Chi1979}  In order to compare the recombination rate to other time constants, it is useful to define  $\tau_{qp-qp}=1/(\Gamma_R n)$.   For thermal quasiparticle populations at 300~mK and 400~mK, $\tau_{qp-qp}=7.7~\mu$s and 1.0~$\mu$s respectively. Recent measurements of quasiparticle lifetime in Al  by Barends et al\cite{barends2009} provide a check to these calculations.   At 210~mK, they measure a lifetime of about 180~$\mu$s and we calculate $\tau_{qp-qp}=210~\mu$s, which is in very good agreement. 

In the thermal model, we separate the total quasiparticle density into two parts, the thermal density $n_{th}$ evaluated at temperature $T_b$ and the excess density $n_{ex}$.  The excess density is an effective increase in the superconductor quasiparticle system temperature due to the power deposited by an NIS junction.  The rate of excess quasiparticle recombination is given by
\begin{equation}
\label{eq_excess_qp_recombination}
\frac{dn_{ex}}{dt} = -\Gamma_R(n_{ex}^2+2n_{ex} n_{th}).
\end{equation}

\begin{figure}\begin{center} 
\includegraphics[width=0.9\linewidth]{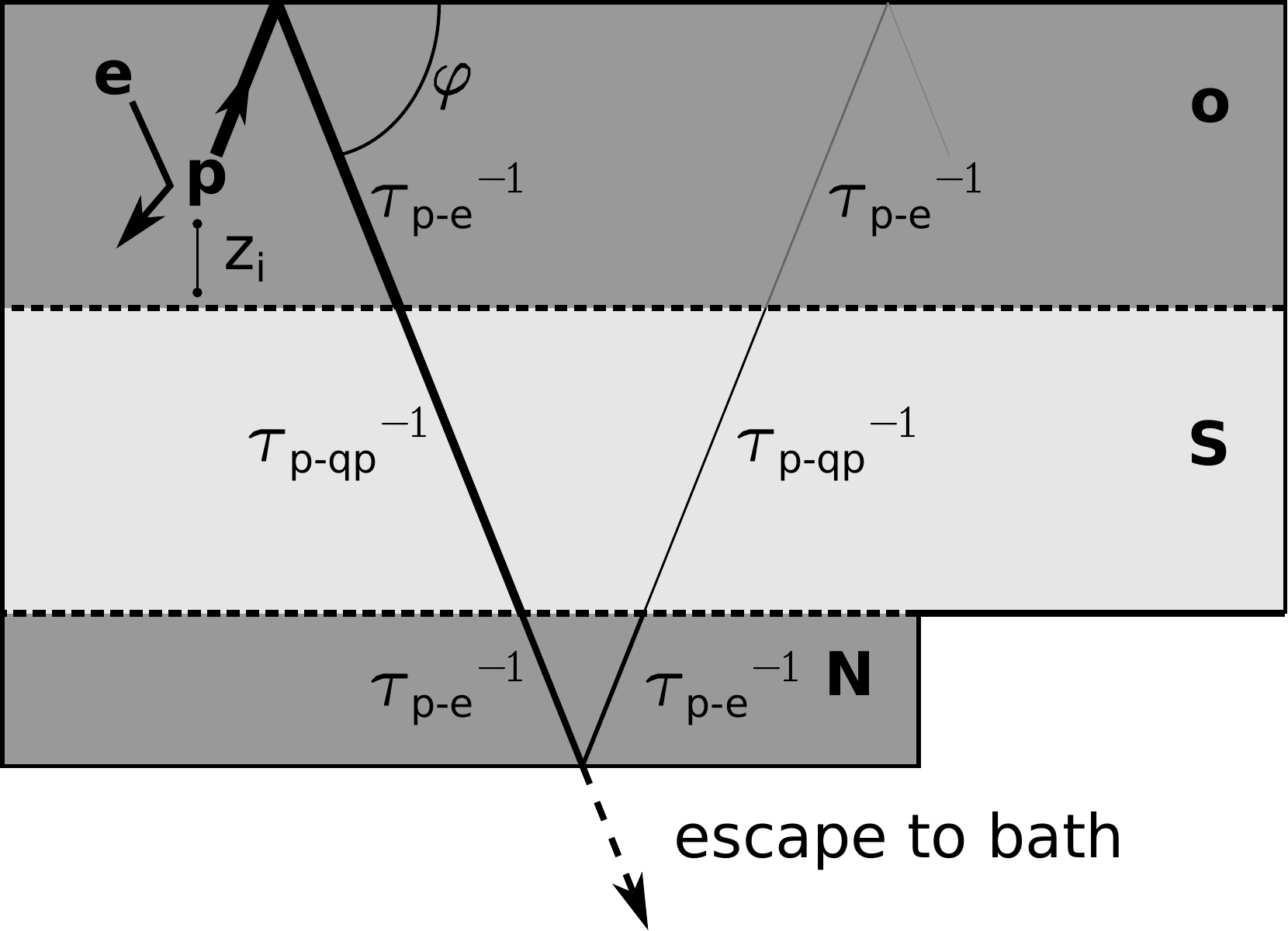}
\caption[Diagram of athermal phonon behavior model.]{\label{fig_phononescapel}The propagation of a phonon starting at position $z_i$ in the $O$ layer and propagating at angle $\varphi$.  The probability amplitude of the phonon is reduced as it propagates through each layer depending on the time constant $\tau$ for the phonon to interact in that layer.  Additionally there is an escape probability $\langle \eta \rangle $ at the interface between the bath and layer $N$.  By following the phonon until the amplitude is reduced to near zero, we calculate the probability that a given phonon either escapes to the bath or is absorbed in the $O$, $S$ or $N$ layer.}
\end{center}\end{figure}

\subsection{Phonon-quasiparticle scattering or ``pair breaking''}
The excitation of electrons by phonons has already been discussed in a normal metal.  A similar process can occur in a superconductor where a phonon of energy $2 \Delta$ or greater destroys a Cooper pair and produces two quasiparticles.  When two quasiparticles recombine, the emitted phonon has sufficient energy to break another Cooper pair.  The pair breaking time for a $2 \Delta$ phonon is given by Kaplan\cite{kaplan1976} and is equal to a material specific parameter $\tau_{p-pb}=230~$ps for Al.  This number is only weakly dependent on temperature and phonon energy.  For example, the rate is only $20\%$ faster for a $3 \Delta$ phonon compared to a $2 \Delta$ phonon.

\section{Phonon escape}
\label{sec:phonon_escape}
Ideally, phonons generated by quasiparticle recombination or electron-phonon scattering escape into the Si substrate, but may instead interact with electrons or quasiparticles before leaving the Al and AlMn layers that make up the NIS refrigerator.  

While the electrons in the $N$, $S$, and $O$ layers are isolated from each other by tunnel barriers, the phonons are not similarly isolated.  All three layers are Al with only small quantities ($\sim4000$~ppma) of Mn in the $N$ and $O$ layers, so phonons should freely pass between these layers.  However, the metal layers making up the NIS refrigerator are deposited on $\sim150~$nm  of SiO$_2$ on a Si wafer 300~$\mu$m thick, and because of this material mismatch the phonons will not always escape into the Si wafer.  The Si wafer is treated as an ideal thermal bath at temperature $T_b$. We model the behavior of  thermal phonons using an acoustic mismatch boundary resistance, and the athermal phonons using a ray tracing model.

\subsection{Thermal phonons: acoustic mismatch}
\label{section_amm}
Phonon thermal resistance at the interface between two different materials arises from reflection due to differences in the speed of sound and is known as acoustic mismatch.  The power flow per unit area from phonons in the combined metal layers at temperature $T_P$ to a SiO$_2$ layer at temperature $T_B$ is 
\begin{equation}
\label{eq:pamm}
\mathcal{P}_{amm} = \xi (T_P^4-T_B^4)
\end{equation}
where $\xi = 360$~pW/($\mu m^2$ K$^4$) for Al on SiO$_2$.\cite{Swartz1987}

\subsection{Thermal phonons: thermal conductivity}
The thermal conductivity $\kappa_P$ of the phonons in the combined metals layers can be calculated as $C_p l_p \langle s\rangle/3$ where $C_p$ is the Debye model heat capacity for phonons, $l_p$ is the mean free path, and $\langle s\rangle$ is the average phonon speed.  Boundary scattering limits the phonon mean free path at low temperatures, so we approximate $l_p=t_S+t_O+t_N$.  Therefore
\begin{equation}
\label{kappa_p}
\kappa_P = \frac{(t_N+t_O+t_S)\langle s\rangle234 N_A n_d k_b}{3} (\frac{T}{\Theta_D})^3
\end{equation}
where the average speed of sound $\langle s\rangle = (s_l +2 s_t)/3 =4.4 \times 10^3$~m/s is a weighted average of the longitudinal $s_l=6.7 \times 10^3$~m/s and transverse $s_t=3.3 \times 10^3$~m/s values for Al.  One could use a different value of $\langle s\rangle$ for phonons originating in the superconductor (primarily longitudinal) or the normal metal (primarily transverse),\cite{Reizer1989, Underwood2011} but the choice of weighting for $\langle s\rangle$ has only a small impact on the results of the thermal model.  The extreme cases of $\langle s\rangle=s_t$ and $\langle s\rangle=s_l$ result in values of $T_N$ that differ by $\sim0.5$~mK, for the parameters in Table \ref{table:central_params_300mK}.  

\subsection{Phonon-phonon scattering}
Phonons can relax by splitting into two phonons of lower energy, but this anharmonic decay process is much slower than other relaxation pathways.  Maris\cite{Maris1990} states that the lifetime for anharmonic phonon decay is of order $\tau_{p-p}\approx 10~\mathrm{ps} (k_b \Theta_D/E)^5$, which is $\sim$10~ms for phonons of energy $2\Delta$ in Al.  There is experimental evidence for phonon mean free paths of order 1~mm in superconducting materials, consistent with long lifetime for anharmonic decay.\cite{Hauser1999} 

\subsection{Athermal phonons}
\label{section_athermalphonons}
Both quasiparticle recombination and electron-phonon scattering by trapped quasiparticles will create athermal phonons. These phonons will interact with electrons or escape before relaxing via anharmonic decay, and therefore their behavior will not be well described by equilibrium physics. In this section, we describe a ray tracing model shown in Fig \ref{fig_phononescapel} for calculating the behavior of these phonons.\footnote{A ray tracing model is not ideal because the phonon wavelengths are of similar size to the metal thin films.  This model is used due to its simplicity and because it results in phonon escape times that scale inversely with the total film thickness, which is consistent with cited work.}  With this model, we calculate the fraction of these phonons which are absorbed in each metal layer, and the fraction which escape to the bath.  The fraction of energy deposited in the $N$ layer is of particular interest because of its direct impact on NIS refrigerator performance. 

Consider a phonon with energy $3\Delta/4$ created at the top of the $O$ layer traveling directly downward with speed of sound $\langle s\rangle$.  The time to propagate from the top of the $O$ layer to the $OS$ interface is $t_O/\langle s\rangle$.  The probability that the phonon scatters with an electron in the $O$ layer in this time is $1-e^{-t_O /(\langle s\rangle\tau_{p-e|3\Delta/4})}$.  If the phonon enters the $S$ layer, the probability that it leaves without interacting is 1 since its energy is below $2\Delta$.  The probability that it interacts in the $N$ layer before reaching the $N$-Bath interface is $1-e^{-t_N /(\langle s\rangle\tau_{p-e|3\Delta/4})}$.  The probability that a phonon incident on the $N$-Bath interface escapes to the bath is $\langle\eta\rangle$.  Therefore, the probability that a phonon created at the top of the $O$ layer escapes to the bath on its first attempt is $e^{-t_O /(\langle s\rangle\tau_{p-e|3\Delta/4})}e^{-t_N /(\langle s\rangle\tau_{p-e|3\Delta/4})}\eta$.  The phonon escape probability $\langle \eta \rangle = (\eta_l +2 \eta_t)/3 =0.71$ is averaged over longitudinal and transverse phonons, and angle.\cite{Ullom1999}  The value of the escape probability has a very weak effect on the results of the thermal model because most phonons are re-absorbed before they have even a single escape attempt.


A phonon is created in either the $O$ layer as shown, or in the $S$ layer at position $z_i$ and with direction given by angle $\varphi$ relative to the junction. We assume specular reflections.   The phonon is given an initial amplitude of 1, which is reduced by a factor $e^{-t/(\tau \langle s\rangle \cos \varphi)}$ in order to propagate the phonon to the next interface, where $t$ is the distance to the interface and $\tau$ is the time constant for the phonon to interact in the current layer.  When the phonon reaches the interface between the $N$ layer and the bath, the amplitude is reduced by a factor $(1-\langle\eta\rangle)$ where $\langle\eta\rangle$ is the transmission probability for a phonon incident on an Al-SiO$_2$ interface.  The probability of a phonon being absorbed in each layer or escaping to the bath is then calculated as the sum of all the amplitude reductions which occur in that layer, averaging over all possible initial values for $z_i$ and $\varphi$. 

The typical time for an electron of energy $\Delta$ to scatter with a phonon is $\tau_{e-p|\Delta}=25$~ns.  The average distance an electron will diffuse in this time is $\sqrt{D_O \tau_{e-p|\Delta} }=13.3~\mu m$, which is much greater than the overlayer trap thickness of 0.5~$\mu m$.  Therefore all values of $z_i$ are equally likely, although it would be necessary to restrict $z_i$ for thicker $O$ layers or greater $\Delta$ because $\tau_{e-p|\Delta} $ falls as $\Delta$ increases.  All values of $\varphi$ are also equally likely.

The variables that result from this calculation are called $A_{E-L}$ where $E$ is replaced with the energy of the phonon and $L$ is replaced with the layer it is associated with.  For example $A_{2\Delta-N}$ is the probability that a $2\Delta$ phonon from recombination is absorbed in the $N$ layer, while $A_{3\Delta/4-O}$ is the probability that a $3/4\Delta$ phonon from quasiparticle trapping is reabsorbed in the $O$ layer.

\begin{table*}[h!b!p!]
\begin{center}
\begin{tabular}{lll}
Input Parameters \\
\hline
Parameter & Input Value & Description \\
\hline
$\mathcal{R}_{NS}$ & 1200~$\Omega \mu m^2$     & junction resistance area product \\
$A_{NS}$                  & 224~$\mu m^2$                     & area of a single refrigerator junction \\
$x_j$                 & 7~$\mu m$                                & length of junction parallel to current flow \\
$x_{edge}$       & 1~$\mu m$                               &  fabrication tolerance, contributes excess $N$ volume \\
$\rho_N$           & 0.0663~$\Omega \mu m$             & resistivity of $N$ layer measured at 4~K \\
$t_N$                & 17.5~nm                                       & thickness pf $N$ layer \\
$d_{side-trap}$ & 3~$\mu m$                              & distance from junction to side-trap \\
$A_{N-extra}$                & 67.5~$\mu m^2$             & parameter to account for excess $N$ layer area \\
$\Sigma$           & 2.3~nW/($\mu m^3$ K$^6$)       & electron-phonon coupling strength \\
$\mathcal{R}_O$ & 60~$\Omega \mu m^2$               &  resistance area product of the overlayer traps \\
$\rho_O$           & 0.0443~$\Omega \mu m$             & resistivity of the $O$ layer  measured at 4~K \\
$t_O$                & 500~nm                                    & thickness of the $O$ layer \\
$x_{end}$       & 1000~$\mu m$                                        & length of overlayer trap \\
$\Delta$           & $190~\mu$eV                                         & energy gap in Al \\
$\tau_{0-qp}$ & 100~ns                                                   & time constant related to quasiparticle recombination rate \\
$\gamma^{-1}$            & 6000                                     & Dynes parameter, adds finite sub-gap states \\
$\rho_{S-normal}$ & 0.00117~$\Omega \mu m$        & resistivity of the $S$ layer in the normal state,  measured at 4~K \\
$t_S$                & 500~nm                                     & thickness of the $S$ layer \\
$T_b$                & 300~mK                                     & bath temperature provided by the cryostat \\
$\xi$                 & 360~pW/($\mu m^2$ K$^4$)               & acoustic mismatch coefficient \\
$\langle \eta \rangle$              & 0.71                                                         & athermal phonon escape probability \\
$\langle s\rangle$                   & 4.4$\times10^9~\mu m$/s              & average phonon speed in Al \\
$N(0)$              & $1.45\times10^{29}~1/($J $\mu m^3)$  & two-spin density of states in Al \\
$E_F$               & 11.63~eV                                               & Fermi energy in Al \\
$\Theta_D$      & 428~K                                                     & Debye temperature in Al \\
$P_{excess}$        & 0~pW                                                       & excess power added to the model \\

\\
Calculated Parameters \\
\hline
Parameter & Calculated Value & Method of Calculation \\
\hline
$R_{NS}$                  & 5.38~$\Omega $                                & $\mathcal{R}_{NS}/A_{NS}$ \\
$x_{side-trap}$   & 10~$\mu m$                              & $x_j+d_{side_trap}$ \\
$A_{N-fridge}$               &   323~$\mu m^2$                     & $(x_j+2.5 x_{edge})(y_j+2 x_{edge})$, area of the $N$ layer \\
$U_N$               & 6.83~$\mu m^3$                          & $(A_N+A_{N-extra})t_N$ \\
$R_{pad}$          & 0.59~$\Omega$		       & $\rho_N (x_j/2+1.5x_{edge})/(t_N y_j)$ \\
$y_j$                 & 32~$\mu m$                              & $A_{NS}/x_j$ \\
$D_{S-normal}$                  & $2.3\times10^{11}~\mu m^2 /s$ & Eq \ref{eq_diffusionconstant_normal} with $\rho_{S-normal}$ \\
$D_{S-I}$                   & $9\times10^9~\mu m^2 /s$          & Eq \ref{eq_qp_diffusion_injection} with $D_{S-normal}$, $T_N$ and $T_{S-left}$   \\
$\Gamma_R$     & 50~$\mu m^3$/s                       & Eq \ref{eq_gamma_r} \\
$\kappa_O$      & $1.66\times10^{-7}~$W/(K $\mu m$) & Eq \ref{eq_weidemannfranz} with $\rho_O$ and $T=300$~mK \\
$\kappa_N$      & $3.70\times10^{-8}~$W/(K $\mu m$) & Eq \ref{eq_weidemannfranz} with $\rho_N$ and $T=100$~mK \\
$\kappa_P$      & $9.83\times10^{-11}~$W/(K $\mu m$) & Eq \ref{kappa_p} with $T=300$~mK \\
$\tau_{p-e|2\Delta}$ & 9.2~ps                                          & Eq \ref{eq_tau_pe_n6} with $E=2\Delta$ \\
$\tau_{p-e|3\Delta/4}$ & 66~ps                                           & Eq \ref{eq_tau_pe_n6} with $E=3\Delta/4$ \\
$\tau_{e-p|\Delta}$ & 38~ns                                                  & Eq \ref{eq_tau_ep} with $E=\Delta$ \\
$\tau_{e-e|\Delta}$ & 540~ns                                              & Eq \ref{tau_ee_clean3d} with $E=\Delta$ \\
$\tau_{tun-OS}$ & 111~ns                                           &Eq \ref{eq_tau_tun} with $t_S$ and $\mathcal{R}_{O}$ \\
$\tau_{tun-SN}$ & 2.2~$\mu$s                                     & Eq \ref{eq_tau_tun} with $t_S$ and $\mathcal{R}_{TN}$ \\
$A_{3\Delta/4-O}$ & 0.86                             & see Appendix \ref{section_athermalphonons} \\
$A_{3\Delta/4-N}$ & 0.02                        & see Appendix \ref{section_athermalphonons} \\
$A_{2\Delta-O}$ & 0.33                                   & see Appendix \ref{section_athermalphonons} \\
$A_{2\Delta-S}$ & 0.38                               & see Appendix \ref{section_athermalphonons} \\
$A_{2\Delta-N}$ & 0.17                                & see Appendix \ref{section_athermalphonons} \\
$\Pi_p$                   &            0.71                                              &  Eq \ref{eq:pi_all} \\ 
$\Pi_{tun}$            &             0.24                                              &  Eq \ref{eq:pi_all} \\
$\Pi_{e}$               &                 0.05                                           & Eq \ref{eq:pi_all} \\
$g$                      &                             -                                 &  if $ x \leq x_j$ then $g(x) = 1$, else $g(x)=0   $ \\
$g_{side-trap}$    &                            -                                &  if $ x \geq x_{side-trap}$ then $g(x) = 1$, else $g(x)=0   $ \\

\hline
\end{tabular}
\caption[Comprehensive thermal model parameters for cooling from 300~mK.]{Numerical values and method used to calculate the parameters in the comprehensive thermal model.  The input parameters are based upon measurements of the properties of the refrigerator described in this paper.  The $N$ layer area depends on the aspect ratio of the tunnel junction due to fabrication tolerances which require a finite amount of AlMn to border the via defined junction.}
\label{table:central_params_300mK}
\end{center}
\end{table*}

\end{document}